\documentclass{article}
\usepackage{spconf,amsmath,graphicx}
\usepackage{hyperref}
\usepackage{bbding} 
\usepackage{multirow}
\usepackage{hyperref} 
\usepackage{bbding} 
\usepackage{multirow}
\usepackage{algorithm}
\usepackage{algorithmic}
\usepackage{colortbl}
\usepackage{cite,xcolor}
\usepackage{diagbox}
\usepackage{booktabs} 
\usepackage{tabularx}
\usepackage{float}
\usepackage{booktabs}
\usepackage{enumitem}

\title{On the effectiveness of enrollment speech augmentation for Target Speaker Extraction
}
\name{Junjie Li$^1$, Ke Zhang$^{2}$, Shuai Wang$^{2,*}$, Haizhou Li$^{2,3}$, Man-Wai MAK$^1$, Kong Aik Lee$^{1,*}$\thanks{$^*$: Corresponding author.} 
\thanks{This work was supported by Project ID P0049192 of the The Hong Kong Polytechnic University, National Natural Science Foundation of China (Grant No. 62401377 and No. 62271432), Shenzhen Science and Technology Program (Grant No. ZDSYS20230626091302006) and Internal Project of Shenzhen Research Institute of Big Data (Grant No. T00120220002 and No. J00220230014).}}
\address{
$^1$ Department of Electrical and Electronic Engineering, The Hong Kong Polytechnic University, Hong Kong \\ 
$^2$ Shenzhen Research Institute of Big Data, Shenzhen, China \\
$^3$ School of Data Science, The Chinese University of Hong Kong, Shenzhen (CUHK-Shenzhen), China 
}

\begin{document}
\ninept 
\maketitle
%
% defination, problem, key contribution
\begin{abstract}
%shuai
Deep learning technologies have significantly advanced the performance of target speaker extraction (TSE) tasks. To enhance the generalization and robustness of these algorithms when training data is insufficient, data augmentation is a commonly adopted technique. Unlike typical data augmentation applied to speech mixtures, this work thoroughly investigates the effectiveness of augmenting the enrollment speech space. We found that for both pretrained and jointly optimized speaker encoders, directly augmenting the enrollment speech leads to consistent performance improvement. In addition to conventional methods such as noise and reverberation addition, we propose a novel augmentation method called self-estimated speech augmentation (SSA). Experimental results on the Libri2Mix test set show that our proposed method can achieve an improvement of up to 2.5 dB.

% Junjie
% Deep learning models have significantly advanced the performance of target speaker extraction (TSE) tasks. However, achieving high performance with deep neural networks (DNNs) requires a large amount of data. Data augmentation is an effective method to generate diverse data, thereby improving the generalization and robustness of DNNs. Unlike previous studies that focus on applying data augmentation methods to the speech mixture, this paper explores applying data augmentations to the enrollment speech, such as adding noise or reverberation to clean enrollment speech, and masking some time or frequency frames in the features of enrollment speech. Additionally, we propose a novel augmentation method called self-estimated speech augmentation (SSA). SSA uses the corresponding estimated speech of the ground truth as the enrollment speech, significantly enhancing performance. Experiments show that our best data augmentation setup can achieve an improvement of up to 2.5 dB.

\end{abstract}
\vspace{-1mm}
\section{Introduction}
\label{sec:intro}
\vspace{-1mm}

% real life human hearing conversation attention 

% TSE definition 

% problems 

% augmentations in other fields 

% 

In real-world environments, people are surrounded by a variety of sounds, such as mixtures of speech from multiple speakers, background noise, and music. The selective attention mechanism \cite{cherry1954some} allows humans to focus on a target speaker's voice while ignoring other interferences in complex noisy scenarios, which is known as the cocktail party problem\cite{bronkhorst2015cocktail}.
Although the attention mechanism in humans is not fully understood, many algorithms have been proposed to emulate  this process in machine learning. Target speaker extraction (TSE) aims to extract the target speech from multi-talker scenarios using audio or visual cues as references \cite{zmolikova2023neural,wang24fa_interspeech} by forming a top-down attention mechanism on the target speaker. TSE serves as a pre-processing step for many speech  downstream tasks, such as automatic speech recognition (ASR) \cite{9383615}, speaker verification \cite{10096848,jin2023speaker}, and text-to-speech (TTS) \cite{yu2024autoprep,botinhao2016speech}.

With the progress of deep learning techniques, deep neural network (DNN)-based methods have achieved remarkable improvements, exemplified by models such as X-Sepformer \cite{liu2023x} and CIENet-mDPTNet \cite{10447529}.
However, a model trained on relatively small datasets lacks reliability and generalizability in real scenarios. Cross-dataset evaluations typically reveal degraded performance and a severe speaker confusion problem, where the interference speaker is extracted instead of the target one \cite{zhao22b_interspeech}. 
% For instance, performance on the WSJ0-2Mix dataset \cite{hershey2016deep,ge20_interspeech} has improved from 15.8 dB to 21.4 dB in terms of SI-SDR,  
 % from SpEx \cite{xu2020spex} to CIENet-mDPTNet \cite{yang2024target}. 

To increase the robustness and generalization, data augmentation is often employed to enhance the  diversity of the training data. 
This approach has been successfully applied to various fields, including speech recognition \cite{park19e_interspeech,ko2017study}, speaker recognition \cite{wang2020data,snyder2018x,yamamoto2019speaker}, speech separation \cite{alex2023data,alex2021mixup,erdogan2018investigations,10094767}, and speech synthesis \cite{bae2024latent}. 
In speech separation tasks \cite{alex2023data,alex2021mixup,erdogan2018investigations,10094767}, it has been  demonstrated that data augmentation can be effectively applied to speech mixtures. These methods can be categorized into \cite{alex2023data} \emph{source-preserving} and \emph{non-source-preserving} augmentations. Source-preserving augmentation only modifies the speech mixture without changing the ground truth sources, while non-source-preserving augmentation modifies both the speech mixture and its ground truth sources. 
% Despite the availability of many large open-source datasets nowadays, data augmentation methods remain crucial in resource-constrained situations.

% \begin{table}[]
% \centering
% \caption{The ResNet34 architecture. \(N\) is the number of speakers. The first and second dimension of the input shows number of filter-banks and the number of time frames. The parameters in structure denote kernel size, channels and stride, respectively. 
% }
% \label{tab:resnet34}
% \begin{tabularx}{\linewidth}{lXc}
% \toprule
% Layer name & Structure & Output \\
% \midrule
% Input & - & $80 \times 200 \times 1$ \\
% Conv2D-1 & $3 \times 3$, $32$, $ 1$ & $80 \times 200 \times 32$ \\
% \midrule
% \multirow{2}{*}{ResNetBlock-1} & $\left[ \begin{array}{c}
% 3 \times 3, 32 \\
% 3 \times 3, 32 \\
% \end{array} \right] \times 3$, $ 1$ & $80 \times 200 \times 32$ \\
% \midrule
% \multirow{2}{*}{ResNetBlock-2} & $\left[ \begin{array}{c}
% 3 \times 3, 64 \\
% 3 \times 3, 64 \\
% \end{array} \right] \times 4$, $ 2$ & $40 \times 100 \times 64$ \\
% \midrule
% \multirow{2}{*}{ResNetBlock-3} & $\left[ \begin{array}{c}
% 3 \times3, 128 \\
% 3 \times 3, 128 \\
% \end{array} \right] \times 6$, $ 2$ & $20 \times 50 \times 128$ \\
% \midrule
% \multirow{2}{*}{ResNetBlock-4} & $\left[ \begin{array}{c}
% 3 \times 3, 256 \\
% 3 \times 3, 256 \\
% \end{array} \right] \times 3$, $ 2$ & $10 \times 25 \times 256$ \\
% \midrule
% StatsPooling & - & $20 \times 256$ \\
% Flatten & - & $5120$ \\
% Dense1 & - & $256$ \\
% Dense2 (Softmax) & - & $N$ \\
% \midrule
% Total & - & - \\
% \bottomrule
% \end{tabularx}
% \vspace{-6mm}
% \end{table}

\begin{figure*}[htbp]
    \centering
    \includegraphics[width=0.85\textwidth]{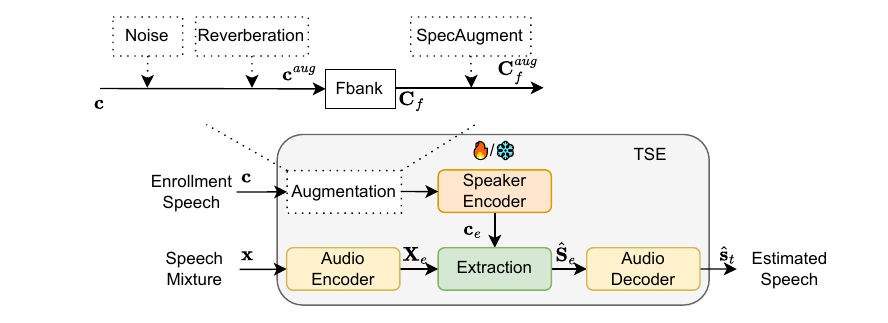}
    \vspace{-6mm}
    \caption{The pipeline of TSE. The speaker encoder could be pretrained from a speaker recognition task and frozen, or jointly trained from scratch with other modules during the training stage. We utilize ResNet34 as the speaker encoder; so the enrollment speech needs to be first transformed to Fbank. Noise and reverberation augmentation methods are applied directly on enrollment speech $\bf c$, and SpecAugment is applied to the Fbank feature $\mathbf{C}_f$.}
    \vspace{-6mm}
    \label{fig:tse}
\end{figure*}

However, there has been limited research on applying augmentations to TSE models. While previous studies on speech separation tasks have proven that augmentations applied to speech mixtures are useful, we propose focusing on augmenting the enrollment speech instead of the speech mixture. A typical TSE model contains two main components 1) a speaker encoder, which extracts the speaker embedding from the enrollment speech, 2) an extraction module, which uses this speaker embedding to extract the target speech from the speech mixture. By applying augmentation to the enrollment speech, we expect the generated speaker embedding to be more diverse, thereby enhancing the robustness of the extraction module and reducing the probability of extracting the wrong speaker.

The contributions of this paper are threefold: First, we propose applying data augmentation on enrollment speech and validate the effectiveness of three common augmentation methods: noise,  reverberation, and SpecAugment \cite{park19e_interspeech}. Second, we propose a novel augmentation method called self-estimated speech augmentation (SSA), which uses the corresponding estimated speech of the target speaker as enrollment speech. Third, we compare the effectiveness of augmentation on enrollment speech versus the speech mixture and find that they can complement each other to some extent, leading to improved overall performance.

\vspace{-1mm}
\section{Methods}
\vspace{-1mm}
\label{sec:method}

\subsection{Formulation of TSE task}
Target speaker extraction  \cite{zmolikova2023neural} aims to isolate the speech signal of a target speaker from a mixture of several speakers using the clues of target speaker. Such clues can be a video of target speaker's lips \cite{li23ja_interspeech,li2024audio}, the direction of target speaker's speech \cite{10508743}, or a pre-recorded enrollment speech \cite{yu2023tspeech,yu23b_interspeech}. In this paper, we focus on single-channel audio-only TSE:
\begin{equation}
\begin{aligned}
\hat{\mathbf{s}}_t &= f_{\text{TSE}}(\mathbf{x},\mathbf{c};\theta),
    \label{equ:tse}
\end{aligned}
\end{equation}
where $\hat{\mathbf{s}}_t$ is the estimated speech  belonging to the same speaker as the enrollment speech $\bf c$. The TSE model $f_{\text{TSE}}(\cdot ;\theta)$ with parameters $\theta$ extracts $\hat{\mathbf{s}}_t$  from the speech mixture:
\begin{equation}
    \mathbf{x} = \mathbf{s}_t+\mathbf{s}_i + \mathbf{n},
\end{equation}
where $\mathbf{s}_t$, $\mathbf{s}_i$, and $\bf n$ are the target, interference,  and noise signals, respectively.

\vspace{-2mm}
\subsection{Pipeline of TSE task}
\label{sec:pipeline}
 A typical TSE system consists of a speaker encoder, an audio encoder, an extraction module, and an audio decoder, as illustrated in Fig. \ref{fig:tse}. The speaker encoder aims to extract the target speaker embedding $\mathbf{c}_e$ from enrollment speech $\bf c$. This speaker embedding, like x-vector \cite{snyder2018x,desplanques2020ecapa} and r-vector \cite{zeinali2019but}, represents the speaker's identity, with each speaker having a unique speaker embedding.
 In TSE tasks,  the speaker encoder could be pre-trained for speaker recognition tasks \cite{wang2018voicefilter, yu2023tspeech}, or jointly trained from scratch with other modules using cross-entropy loss \cite{xu2020spex,ge20_interspeech}. 
 The audio encoder transforms the speech mixture signal $\bf x$ into the speech mixture embedding $\mathbf{X}_e$, while the audio decoder renders the estimated speech $\hat{\mathbf{s}}_t$ from its latent latent representation $\hat{\mathbf{S}}_e$. The extraction module aims to estimate the target speech representation $\hat{\mathbf{S}}_e$ from the speech mixture embedding $\mathbf{X}_e$, using the speaker embedding $\mathbf{c}_e$ as a reference. 

\noindent \textbf{ResNet34:} 
In this paper, we utilize the ResNet34 architecture defined in Wespeaker toolkit\footnote{\url{https://github.com/wenet-e2e/wespeaker/blob/master/wespeaker/models/resnet.py}} as the speaker encoder \cite{zeinali2019but,wang2023wespeaker}. ResNet34 takes 2-dimensional Fbank features as input and processes them using 2-dimensional CNN layers. The output is then passed through four ResNet blocks. In this paper, we consider both pretrained and jointly trained speaker encoders. For the pretrained setup, the  checkpoint from Wespeaker\footnote{\url{https://wespeaker-1256283475.cos.ap-shanghai.myqcloud.com/models/voxceleb/voxceleb_resnet34_LM.zip}} is utilized. This checkpoint is trained using the VoxCeleb2 dataset \cite{chung2018voxceleb2} with several data augmentation techniques, such as adding noise and reverberation. Detailed training procedures are provided in \cite{wang2023wespeaker}. The pretrained speaker model achieves Equal Error Rates (EER) of 0.797\%, 0.937\%, and 1.695\% on the Vox1-O-Clean, Vox1-E-Clean, and Vox1-H-Clean subsets, respectively\footnote{\url{https://github.com/wenet-e2e/wespeaker/blob/master/examples/voxceleb/v2/README.md}}. For the jointly trained setup, we only load the definition of the ResNet34 architecture and initialize it randomly.

\noindent \textbf{BSRNN}:
For the extraction module, we adopt the band-split RNN (BSRNN)~\cite{luo2023music}, which has been proved to be effective in music source separation and personal speech enhancement tasks \cite{luo2023music,yu23b_interspeech,yu2023tspeech}. 
BSRNN comprises three sub-modules: the \emph{band split module}, the \emph{band and sequence modeling module}, and the \emph{mask estimation module}. The band split module divides the fullband complex-valued spectrogram $\mathbf{X}_e$ of the speech mixture into several subband spectrograms. Each subband spectrogram is processed through a layer normalization and a fully connected layer, then the outputs are merged to generate a new real-valued feature. In the case of the TSE task, the speaker embedding \(\mathbf{c}_e\) is replicated to align with the time dimension of the real-valued  feature.
The repeated speaker embeddings and the real-valued feature are then element-wise multiplied to generate a fused feature, which serves as the input to the band and sequence modeling module. The band and sequence modeling module is similar to the dual-path RNN architecture \cite{luo2020dual}, designed to perform interleaved sequence-level and band-level modeling via two different residual RNN layers. The mask estimation module splits its input into subband features, estimates a mask for each subband feature and merge these subband masks into a full mask.  The full mask is then element-wise multiplied with the output of the band split module to generate the estimated  spectrogram \( \hat{\mathbf{S}}_e \).

The audio encoder and decoder are implemented using short-time Fourier transform (STFT) and inverse STFT, respectively.

\vspace{-2mm}
\subsection{Loss function}
A scale-invariant signal-to-distortion ratio (SI-SDR) \cite{le2019sdr} loss is used to measure the quality between the estimated and clean target speech:
\begin{equation}
    \mathcal{L}_{\text{SI-SDR}}(\mathbf{s}_t,\hat{\mathbf{s}}_t) = -10 \log_{10}\frac{||\frac{<\hat{\mathbf{s}}_t,\mathbf{s}_t>\mathbf{s}_t}{||\mathbf{s}_t||^2}||^2}{||\hat{\mathbf{s}}_t-\frac{<\hat{\mathbf{s}}_t,\mathbf{s}_t>\mathbf{s}_t}{||\mathbf{s}_t||^2}||^2}.
\end{equation}
 When the speaker encoder is learnable, we add a cross-entropy (CE) loss for speaker classification:
\begin{equation}
    \mathcal{L}_{\text{CE}}(\mathbf{c}) = -\sum^N_{n=1}\mathbf{y}_n \log(\operatorname{softmax}(\mathbf{W}  \mathbf{c}_e)).
\end{equation}
The final loss is given by:
\begin{equation}
    \mathcal{L} = (1-\gamma)\mathcal{L}_{\text{SI-SDR}}(\mathbf{s}_t,\hat{\mathbf{s}}_t)+\gamma \mathcal{L}_{\text{CE}}(\mathbf{c}),
\end{equation}
where  $\mathbf{s}_t$ and  $\mathbf{y}_n$  are clean target speech and target speaker’s class label, respectively. $N$ is the number
of speakers in the dataset. $\bf W$
is a learnable weight matrix  for speaker classification. \(\gamma\) is a scaling factor, while \(\mathbf{c}_e\) refers to the speaker embedding of the enrollment speech \(\bf c\). 
As shown in Fig. \ref{fig:tse}, when the speaker encoder is frozen, we set \(\gamma\) to 0; otherwise, we set \(\gamma\) to 0.1.

\vspace{-1mm}
\section{Augmentations on enrollment speech}
\vspace{-1mm}
\label{sec:aug}
Data augmentation has been investigated for many years and is an effective way to improve the robustness of DNNs. In this section, we propose applying data augmentation strategies to enrollment speech, which shows significant improvement.
 We first adopt three common data augmentation methods: noise, reverberation, and SpecAugment \cite{park19e_interspeech}, as illustrated in Fig. \ref{fig:tse}. Additionally, we propose a new data augmentation method, self-estimated speech augmentation (SSA)\footnote{This will be released in \url{https://github.com/wenet-e2e/wesep}.}, which uses the estimated speech as enrollment speech. 

\vspace{-3mm}

\subsection{Noise}
We use non-speech noises from the MUSAN corpus \cite{snyder2015musan} to generate an augmented enrollment speech with  probability $\beta$. Otherwise, the clean signal $\bf c$ is used as enrollment speech:
\begin{equation}
\label{equ:noise}
\begin{aligned}
\mathbf{c}^{aug} &=
\begin{cases} 
    \mathbf{c} + \alpha\mathbf{ n} & \text{if } \mathcal{U}(0,1) < \beta \\ 
    \bf c & \text{otherwise}.
\end{cases} 
\end{aligned}
\end{equation}
Here, $\mathbf{c}^{aug}$ denotes the augmented enrollment speech, where $\alpha$ controls the amplitude factor of the noise signal $\bf n$:
\begin{equation}
    \alpha = 10^{-\frac{\text{SNR}}{20}}.
\end{equation}
The signal-to-noise ratio (SNR) \cite{vincent2006performance} is randomly sampled between $-5$ and $15$ dB: 
\begin{equation}
    \text{SNR} \sim \mathcal{U}(-5, 15),
\end{equation}
where $\mathcal{U}$ is a uniform distribution. 
\vspace{-1mm}
\subsection{Reverberation}
Unlike the commonly adopted approach where the reverberation is sampled from datasets such as RIRNoise, we use FRAM-RIR \cite{luo2023fast} to generate room impulse response (RIR) filters in an online manner. FRAM-RIR is able to simulate realistic RIR filters in a fast and efficient way. The
reverberation time $T_{60}$ is randomly sampled within $[0.1, 0.7]$
seconds. The room size is sampled with dimensions varying
from $3 \times 3 \times 2.5$ to $10 \times 10\times 4\text{ } m^3$ (length$\times$width$\times$ height).
Like the noise augmentation, we generate the augmented speech at probability $\beta$:
\begin{equation}
\label{equ:reverb}
\begin{aligned}
\mathbf{c}^{aug} &=
\begin{cases} 
    \mathbf{c} * h & \text{if } \mathcal{U}(0,1) < \beta \\ 
    \bf c & \text{otherwise},
\end{cases}
\end{aligned}
\end{equation}
where $h$ denotes the simulated RIR filter and $*$ denotes convolution.

\vspace{-1mm}
\subsection{SpecAugment}
For the ResNet34 structure, the input $\bf c$ needs to be first transformed into a Fbank feature $\mathbf{C}_{f} \in \mathcal{R} ^{T\times F}$, where $T$ and $F$ denote  the number of frames and frequency bins, respectively. SpecAugment \cite{park19e_interspeech} is originally designed for ASR tasks, applying time and frequency masks on speech features:
\begin{equation}
\label{equ:specaug}
\mathbf{C}^{aug}_{f} =
\begin{cases} 
    \mathbf{C}_{f} \odot \mathbf{ M} & \text{if } \mathcal{U}(0,1) < \beta \\ 
    \mathbf{C}_{f} & \text{otherwise},
\end{cases} \\
\end{equation}
where $\mathbf{C}_{f}$ and $\mathbf{C}^{aug}_f$ represents original Fbank and augmented Fbank features, respectively. $\odot$ denotes the element-wise multiplication. $\bf M$ is a masking matrix which has the same shape with $\mathbf{C}_{f}$:
\begin{equation}
\begin{aligned}
    \mathbf{M}[t,f] &= 
\begin{cases} 
    0 & \text{if } t \in[t_s,t_s+t_l] \text{ or } f \in[f_s,f_s+f_l] \\ 
    1 & \text{otherwise},
\end{cases} 
\end{aligned}
\end{equation}
where $t_s$, $t_l$, $f_s$, and $f_l$ denote the start frame and length of masking in the time and frequency domains, respectively:
\begin{equation}
    \begin{aligned}
        t_s &\sim \lfloor \mathcal{U}(0,T) \rfloor\\
t_l &\sim \lfloor \mathcal{U}(0,11) \rfloor \\
f_s &\sim \lfloor \mathcal{U}(0,F) \rfloor \\
f_l &\sim \lfloor \mathcal{U}(0,9)  \rfloor \\
    \end{aligned}
\end{equation}
where $\lfloor\cdot \rfloor$ denotes the floor of an integer.

\vspace{-1mm}
\subsection{Self-estimated speech augmentation}
\begin{figure}[htbp]
    \centering
    \includegraphics[width=0.5\textwidth]{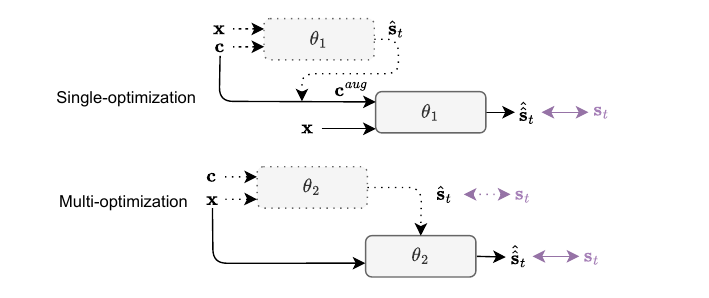}
    \vspace{-6mm}
    \caption{Within the single-optimization method, the TSE models share the same parameters. Similarly, within the multi-optimization method, the TSE models also share the same parameters. }
    \label{fig:ssa}
    \vspace{-3mm}
\end{figure}

To expand the diversity of training data, the augmented speech should be significantly different from the original clean speech in terms of signal while preserving the speaker's identity. Following this rationale, we propose utilizing the estimated speech as the enrollment speech, a method we call self-estimated speech augmentation (SSA). Although the estimated speech and the enrollment speech share the same speaker identity, their signals are completely different not only because of the different content but also due to the introduction of signal artifacts. Such distortion is significantly different from those introduced by the previously mentioned methods.
% Previous methods, such as adding noise or reverberation, can increase data diversity, but the speaker's speech signal remains unchanged. The SpecAugment method alters speech features by masking some frames, but the frames that are not masked remain the same as before. 
In this section, we introduce two ways to implement SSA: single-optimization and multi-optimization, as shown in Fig. \ref{fig:ssa}. 

\noindent \textbf{Single-optimization:} 
Similar to the methods introduced above, we still rely on probability $\beta$ to decide whether to augment the enrollment speech: 
\begin{equation}
\label{equ:ssa11}
\begin{aligned}
\mathbf{c}^{aug} &=
\begin{cases} 
    \hat{\mathbf{s}}_t & \text{if } \mathcal{U}(0,1) < \beta \\ 
    \bf c & \text{otherwise},
\end{cases} 
\end{aligned} 
\end{equation}
where $\hat{\mathbf{s}}_t$ is the estimated speech from the TSE model with original enrollment speech $\bf c$ as input: 
\begin{equation}
    \hat{\mathbf{s}}_t = f_{\text{TSE}}(\mathbf{x},\mathbf{c};\theta).
\end{equation}
With the augmented enrollment speech $\mathbf{c}^{aug}$, the TSE model is guided to predict the final target speech $\hat{\hat{\mathbf{s}}}_t$: 
\begin{equation}
\begin{aligned}
    \hat{\hat{\mathbf{s}}}_t &= f_{\text{TSE}}(\mathbf{x},\mathbf{c}^{aug};\theta).
\end{aligned} 
\end{equation}
And the final loss is given by:
\begin{equation}
    \mathcal{L} = (1-\gamma) \mathcal{L}_{\text{SI-SDR}}(\mathbf{s}_t,\hat{\hat{\mathbf{s}}}_t)+\gamma \mathcal{L}_{\text{CE}}(\mathbf{c}^{aug}).
\end{equation}

\noindent \textbf{Multi-optimization}: 
Instead of using different types of enrollment speech in separate training batches, we propose to use various types of enrollment speech within the same training batch. And the final loss is given by:
\begin{equation}
\mathcal{L} = (1-\beta) \mathcal{L}_{1} + \beta \mathcal{L}_2,
\end{equation}
where $\beta$ is originally used to control the probability of adding augmentations in  Equation \ref{equ:noise}. To ensure fairness between multi-optimization and single-optimization, we use $\beta$ to control the weight of loss here. And $\mathcal{L}_1$ is used to optimize $\hat{\mathbf{s}_t}$: 
\begin{equation}
    \mathcal{L}_1 = (1-\gamma)\mathcal{L}_{\text{SI-SDR}}(\mathbf{s}_t,\hat{\mathbf{s}}_t)+\gamma \mathcal{L}_{\text{CE}}(\bf c).
\end{equation}
$\mathcal{L}_2$ is used to optimize $\hat{\hat{\mathbf{s}}}_t$:
 \begin{equation}
     \mathcal{L}_2= (1-\gamma)\mathcal{L}_{\text{SI-SDR}}(\mathbf{s}_t,\hat{\hat{\mathbf{s}}}_t)+\gamma \mathcal{L}_{\text{CE}}(\hat{\mathbf{s}}_t).
 \end{equation}

\vspace{-5mm}
\section{Experimental details}
\vspace{-1mm}
\label{sec:exp}
\subsection{Datasets}
\textbf{Training data:}
We conducted experiments using two versions of  Libri2mix~\cite{libri2mix} dataset: `Mix\_clean' and `Mix\_both'. Both Mix\_clean and Mix\_both consist of mixtures of two speakers. The difference is that  Mix\_both contains additional noise signals from WHAM! dataset \cite{wichern2019wham}. Each version contains four types of sets: train-360, train-100, dev and test. There are no ovelapping utterances or speakers in each set. And we  combined train-360 and train-100 together to obtain a new set train-460.  The detailed statistics are shown in Table \ref{tab:dataset}. 
In the training stage, the enrollment speech of the target speaker was randomly sampled.  The speech mixture was randomly chunked into 3-second segments, while the entire utterance remained to be the enrollment speech.

\noindent \textbf{Evaluation data:} We utilized test set of Mix\_both and Mix\_clean from Libri2mix as in-domain evaluation. The enrollment of target speaker in dev and test sets comes from TD-SpeakerBeam script \footnote{\url{https://github.com/BUTSpeechFIT/speakerbeam}}.  
English  WSJ0-2mix \footnote{\url{https://github.com/gemengtju/SpEx_Plus/tree/master/data/wsj0_2mix}} \cite{ge20_interspeech} and Mandarin Aishell2Mix \footnote{\url{https://github.com/jyhan03/icassp22-dataset}} \cite{han2022dpccn} test set were  regarded as out-domain evaluation sets. 

For all training and evaluation datasets, fully overlap (min. version) and 16kHz sampling rate were configured.

\renewcommand{\arraystretch}{0.9}
\begin{table}[htbp]
    \centering
    \caption{Statistics of Libri2mix.}
    \begin{tabular}{c|c|c|c|c}
    \midrule
         \multirow{3}{*}{Mix\_both} &  Sets & Utterances & Hours & Speakers \\ \cmidrule{2-5}
         & train-460 & 64,700& 270 & 1172 \\
         or &  train-360 & 50,800 & 212 &921 \\ 
          \multirow{2}{*}{ Mix\_clean} & train-100  & 13,900 & 58 & 251 \\
         & dev & 3,000& 11 & 40 \\
         & test & 3,000 & 11 &40 \\
         \midrule
    \end{tabular}
    \vspace{-6mm}
    \label{tab:dataset}
\end{table}

\vspace{-1mm}
\subsection{Experimental settings}
We set $\beta$ to 0.6\footnote{We refer to the parameters in Wespeaker to determine this probability.}. 
 All models  were trained for 100 epochs on segments with 3 seconds.
 The initial learning rate  and final learning rate are  10e-3 and 2.5e-5, respectively. And current learning rate decays according to:
 \begin{equation}
\text{current}_{\text{lr}}=\text{initial}_{\text{lr}} \times \exp( \frac{\text{current}_{\text{iter}}}{\text{max}_{\text{iter}}}\times \log_{e} \frac{\text{final}_{\text{lr}}}{\text{initial}_{\text{lr}}}
 ),
 \end{equation}
where current$_{\text{lr}}$, initial$_{\text{lr}}$, and final$_{\text{lr}}$ denote the current, the initial, and the final learning rate, respectively. current$_{\text{iter}}$ and max$_{\text{iter}}$ denote the current and the max iteration during the training stage, respectively. 

We averaged the last 5 training checkpoints and regard it as the best one for inference. 

\vspace{-1mm}
\subsection{Evaluation metrics}
We utilize SI-SDR, signal-to-distortion ratio (SDR) \cite{le2019sdr}, and accuracy (Acc.) to evaluate different systems. Accuracy measures the proportion of test samples with an SI-SDR improvement (SI-SDRi) above 1 dB; otherwise we consider the TSE system extracted the incorrect speaker or outputted the speech mixture. 

\renewcommand{\arraystretch}{0.9}
\begin{table*}[t]
	\centering
	\caption{Systems were trained on Mix\_both dataset and evaluated on Mix\_both and Mix\_clean test sets in terms of SI-SDR (dB) and accuracy (\%). `S' and `M' denote `Single-optimization' and `Multi-optimization', respectively. The `Pretrained speaker encoder' means the speaker encoder is pretrained from speaker recognition tasks and remains frozen during the TSE training stage. $\uparrow$ means higher values are better.  }
	\label{tab:noise}
	\begin{tabular}{c|c|c|c|c|c|c|c|c|c|c|c}
		\midrule
		  \multirow{3}{*}{Sys. \#}& Training & Pretrained &\multicolumn{5}{c|}{Augmentation}     & \multicolumn{2}{c|}{Mix\_both} &\multicolumn{2}{c}{Mix\_clean}   \\ \cmidrule{4-8} \cmidrule{9-12}
     &set &  speaker& \multirow{2}{*}{Noise} &\multirow{2}{*}{Reverberation} & \multirow{2}{*}{SpecAugment}& \multicolumn{2}{c|}{SSA} & \multirow{2}{*}{SI-SDR$\uparrow$ }  &\multirow{2}{*}{Acc.$\uparrow$}  &  \multirow{2}{*}{SI-SDR$\uparrow$} &\multirow{2}{*}{Acc.$\uparrow$	 } \\ \cmidrule{7-8}
     & &encoder &  & & &S &M  & & & &      \\ \midrule 
     Mixture &\multicolumn{7}{c|}{} &-1.94 &- & 0.00& -     \\ \midrule 
     1& \multirow{10}{*}{Train-} & \multirow{8}{*}{\Checkmark} & & & & & & 6.11&86.78 & 9.81 &88.47  \\  \cmidrule{1-1} \cmidrule{4-8} \cmidrule{9-12}
     2 & \multirow{9}{*}{100} & & \Checkmark& & & & & 6.88 &89.12 &10.46 & 90.72  \\ \cmidrule{1-1} \cmidrule{4-12} 
     3 & &  & & \Checkmark & & &  &6.72 &88.47 &10.22 & 89.43  \\  \cmidrule{1-1} \cmidrule{4-12} 
     4 & & & & & \Checkmark  & &  &6.57 &88.07  & 9.79 &88.50  \\  \cmidrule{1-1} \cmidrule{4-12} 
     5& & & &  & & \Checkmark & &  7.50&90.53  &11.13 &91.57   \\  \cmidrule{1-1} \cmidrule{4-12} 
     6 & & & & & &  &\Checkmark &  8.24&91.22 & 12.12 &92.07 \\ \cmidrule{1-1} \cmidrule{4-12} 
     7 & & & \Checkmark& \Checkmark &\Checkmark &  &\Checkmark &  8.11& \textbf{92.72}  &11.78 & \textbf{93.65}  \\  \cmidrule{1-1} \cmidrule{4-12} 
     8 & & &  \Checkmark& \Checkmark & &  &\Checkmark &  \textbf{8.64} & 92.60&\textbf{12.51}& 93.52   \\ \cmidrule{1-1} \cmidrule{3-12}  
     
        9 & & \multirow{6}{*}{\XSolidBrush}& & & & & & 8.20& 93.57& 11.61 &93.78  \\  \cmidrule{1-1} \cmidrule{4-12} 
        10& & & \Checkmark & & & & &  8.72&95.52 &12.13 & 95.55 \\ \cmidrule{1-1} \cmidrule{4-12} 
        11 & & &  &\Checkmark  & & & &  8.35&93.93 &11.68 &93.83  \\ \cmidrule{1-1} \cmidrule{4-12} 
        12 & & &  &  &\Checkmark & &  &8.24 &94.48 &11.51 &94.23   \\ \cmidrule{1-1} \cmidrule{4-12} 
        13 & & &  &  & &\Checkmark & &8.99 &94.2 & 12.63&95.33   \\ \cmidrule{1-1} \cmidrule{4-12} 
        
        14 & &  &  &  & & &\Checkmark &  \textbf{9.82}& \textbf{96.28} & \textbf{13.74} & \textbf{96.40} \\ \cmidrule{1-1} \cmidrule{4-12} 
       15 & &  & \Checkmark& \Checkmark & &  &\Checkmark &  9.25 & 95.17 &12.98&95.73  \\    \midrule
       % \cmidrule{1-1} \cmidrule{4-12} 
       % 16 initM  & & & \Checkmark& \Checkmark & &  &\Checkmark &  9.79& \textbf{96.43} & 13.58 &96.35 \\   \cmidrule{1-1} \cmidrule{4-12} 
       % 17 init  &  & & \Checkmark& \Checkmark & &  &\Checkmark &  9.01& 95.37&12.62 & 95.80 \\ \midrule 
        16 & \multirow{3}{*}{Train-} &\multirow{3}{*}{\Checkmark} & & & & & & 10.81& 97.12 & 14.96 &97.52   \\  \cmidrule{1-1} \cmidrule{4-12} 

        17 &\multirow{2}{*}{460} & & \Checkmark& \Checkmark & &  &\Checkmark & \textbf{11.04}& \textbf{97.17} & \textbf{15.32 } &\textbf{97.72}  \\   \cmidrule{1-1} \cmidrule{3-12}
        - & & \multicolumn{6}{c|}{Espnet TD-SpeakerBeam \footnotemark } & 10.74 & - &-  &-\\ \midrule
			\end{tabular} 
   \vspace{-6mm}
\end{table*}
\footnotetext{ \url{https://github.com/espnet/espnet/blob/master/egs2/librimix/tse1/README.md}}

\renewcommand{\arraystretch}{0.9}
\begin{table}[]
    \centering
    \caption{The equal error rate (EER) (\%) of the speaker encoder across different systems. $\downarrow$ means lower values are better.}
    \begin{tabular}{c|c|c|c|c|c|c|c}
        \midrule
        Sys. \# &1$\sim$8 &9 &10 &11 &12 &13 &14   \\ \midrule
        Pretrained & \multirow{3}{*}{\Checkmark} &\multicolumn{6}{c}{\multirow{3}{*}{\XSolidBrush}}  \\
        speaker & & \multicolumn{6}{c}{}  \\ 
        encoder &  &  \multicolumn{6}{c}{}  \\ \midrule
        EER$\downarrow$ &0.82  &6.83 &7.23 &7.23 & 6.77 &7.95 & 6.95 \\ \midrule 
    \end{tabular}
    \vspace{-6mm}
    \label{tab:EER}
\end{table}

\renewcommand{\arraystretch}{0.9}
\begin{table*}[t]
	\centering
	\caption{Systems are trained on the Mix\_clean train-100 set.  `Aug.' denotes the utilization of the integrated augmentation, like System 8 and 15. `CL' means we first train the model without any augmentation for 50 epochs, followed by augmentation for the next 50 training epochs.}
	\label{tab:100_clean}
	\begin{tabular}{c|c|c|c|c|c|c|c|c|c|c|c}
		\midrule
		  \multirow{3}{*}{Sys. \#}&Pretrained & \multirow{3}{*}{Aug.}& 
    \multicolumn{5}{c|}{Libri2mix}& \multicolumn{2}{c|}{WSJ0-2mix} & \multicolumn{2}{c}{Aishell2Mix} \\ \cmidrule{4-12} 
    &speaker & & \multicolumn{2}{c|}{Mix\_both} &\multicolumn{3}{c|}{ Mix\_clean} & \multirow{2}{*}{SI-SDR$\uparrow$}& \multirow{2}{*}{Acc.$\uparrow$} & \multirow{2}{*}{SI-SDR$\uparrow$} & \multirow{2}{*}{Acc.$\uparrow$}\\ \cmidrule{4-8} 
     & encoder & &  SI-SDR$\uparrow$  & Acc.$\uparrow$  & SI-SDR 
 $\uparrow$ &SDR$\uparrow$ & Acc.$\uparrow$& & & &   \\ \midrule
     Mixture & &   & -1.94& - & 0.00 &- &- & 0.00 &- & 2.50 &- \\ \midrule
     18&\multirow{2}{*}{\Checkmark} &  \XSolidBrush &3.88 &83.58 & 12.56 & 13.05&91.72&  14.43 & 95.72 &1.22 &51.67 \\  
    19 & & \Checkmark &\textbf{4.20}  &\textbf{83.93} &\textbf{13.68} &\textbf{14.03} &\textbf{94.48}&\textbf{14.71} &\textbf{97.32}&\textbf{4.58} & \textbf{62.60}  \\ \midrule
     20& \multirow{3}{*}{\XSolidBrush}& \XSolidBrush & 4.20& 84.02 &  14.31   &15.06 & 95.12 &14.65  &96.45 & 7.16 &82.70 \\  
     21 & & \Checkmark & 3.90 & \textbf{84.27} &   14.22  &14.85  & \textbf{95.88}  &14.93 &\textbf{98.45} & 7.53 &85.17 \\
     22 (CL) \footnotemark &  & \Checkmark & \textbf{4.26}& 84.23 &   \textbf{14.57}   &\textbf{15.23} &  95.83&\textbf{15.19} &98.02  & \textbf{7.70} & \textbf{86.30} \\  
          % 22  &  & \Checkmark & \textbf{4.91}& 85.88 & 14.62 & & 96.08 &  15.01 & 97.52 & 8.17 &87.27  \\  
          % 22 (initM)  &  & \Checkmark & \textbf{4.89}& \textbf{87.05} & 14.55 & 15.18& \textbf{96.27} &  15.04 & 98.05 &8.07 &88.37  \\ 
     \midrule
    \multicolumn{3}{c|}{SSL-TD-SpeakerBeam \cite{10448315}} &-& -&14.65&15.26 & 97.0 & - & - & - & - 
 \\\midrule 
 \multicolumn{3}{c|}{TD-SpeakerBeam \cite{9054683,10448315}} &-& - &13.03 &13.69 & 95.2 & - & - & - & - \\\midrule 
 \multicolumn{3}{c|}{Target-Confusion \cite{zhao22b_interspeech}}  &-& -&13.88 &- & - & - & - & - & - \\\midrule
 \multicolumn{3}{c|}{sDPCCN \cite{han2022dpccn}}  &-& -&11.61 &- & - & - & - & 5.78 & - \\\midrule
  \multicolumn{3}{c|}{SpEx+ \cite{chen2023mc,ge20_interspeech}}  &-& -&13.41 &- & - & - & - & - & - \\\midrule

  \multicolumn{3}{c|}{MC-SpEx \cite{chen2023mc}}  &-& -&14.61 &- & - & - & - & - & - \\\midrule
 
			\end{tabular} 
   \vspace{-6mm}
\end{table*}
\footnotetext{We have also tried the augmentation setup of  System 14, but its performance is worse than this.}

\renewcommand{\arraystretch}{0.9}
\begin{table*}[h]
    \centering
    \caption{Systems are trained on the Mix\_clean train-100 set. 
\textbf{Only noise augmentation} is applied in these experiments}
    \begin{tabular}{c|c|c|c|c|c|c|c|c|c|c|c}
    \midrule
         \multirow{3}{*}{Sys. \#} & Pretrained & Enrollment& Speech & \multicolumn{4}{c|}{Libri2mix} & \multicolumn{2}{c|}{WSJ0-2mix} & \multicolumn{2}{c}{Aishell2Mix} \\ \cmidrule{5-12} 
         & speaker& speech& mixture& \multicolumn{2}{c|}{Mix\_both}&\multicolumn{2}{c|}{ Mix\_clean} & \multirow{2}{*}{SI-SDR$\uparrow$}& \multirow{2}{*}{Acc.$\uparrow$} &  \multirow{2}{*}{SI-SDR$\uparrow$} &  \multirow{2}{*}{Acc.$\uparrow$} \\ \cmidrule{5-8} 
         &encoder& & & SI-SDR$\uparrow$ & Acc.$\uparrow$ & SI-SDR$\uparrow$ & Acc.$\uparrow$ & & & &   \\ \midrule
          Mixture& & &  & -1.94 & - & 0.00 & - & 0.00 & - & 2.50 & - \\ \midrule  
         18 & \multirow{4}{*}{\Checkmark} & & &3.88 & 83.58&12.56&91.72 & 14.43 &95.72&1.22 & 51.67\\ 
         23 & & \Checkmark & &3.38 &81.38 & \textbf{12.66} & 91.28 &  \textbf{14.70} &\textbf{96.45}&  \textbf{3.22} &\textbf{59.30} \\
         24 & & & \Checkmark &6.87 &88.47 & 11.86 &90.90 &13.77 & 94.87 & -0.01 & 50.97 \\
         25 & & \Checkmark & \Checkmark  &   \textbf{7.39} &\textbf{89.97} & 12.13& \textbf{91.52}& 14.08&96.08 & 1.62 & 56.33\\ \midrule
         20&\multirow{4}{*}{\XSolidBrush} & & & 4.20 & 84.02 & \textbf{14.31} & 95.12 &  14.65 &96.45 &  7.16 & 82.70 \\ 
         26 (CL) & &  \Checkmark & & 4.41 &85.57 &  14.26 &95.60 &  \textbf{14.97} &97.88& \textbf{7.40}& 84.57 \\ 
         27 (CL) & & & \Checkmark &8.68 & 93.80 & 14.22 & 95.80 & 14.52& 96.92  &6.59 & 84.03 \\
         28 (CL) & &  \Checkmark & \Checkmark &\textbf{8.72} &\textbf{94.27} &  14.18& \textbf{95.95} &  14.73 & \textbf{97.93} &6.89 & \textbf{85.50}  \\ \midrule  
    \end{tabular}
    \vspace{-6mm}
    \label{tab:kind}
\end{table*}

\vspace{-1mm}
\section{Results and Analysis}
\vspace{-1mm}
\label{sec:result}

\vspace{-1mm}
\subsection{Comparative analysis of data augmentation }
\label{sec:noise}

Table \ref{tab:noise} shows the performance of systems trained on Mix\_both version using different data augmentation methods. From this table, we can draw the following conclusions:

\begin{enumerate}[leftmargin=0.4cm]
    \item For both pretrained and jointly-trained speaker encoders, all augmentation methods enhance performance. SpecAugment shows the least improvement and even slight degradation in the Mix\_clean test set regarding SI-SDR. In contrast, our proposed SSA method yields the most significant improvement. Furthermore, within SSA, multi-optimization outperforms single-optimization. 
    \item Generally, the jointly-optimized speaker encoder outperforms its pretrained counterpart, as evidenced by a 2 dB difference in SI-SDR between System 1 and System 9. This likely occurs because the pretrained encoder is over-optimized towards the speaker recognition task and tends to produce similar embeddings for different utterances from the same speaker, reducing input diversity for the extraction module. 
    % Consequently, the extraction module struggles to generalize effectively, leading to decreased performance. 
    \item We observe that SpecAugment negatively impacts SI-SDR but improves accuracy, as seen in the comparison between System 7 and System 8. The literature \cite{wang2023wespeaker} indicates SpecAugment is usually incompatible with other augmentations, so we neglect it in our final integrated approach (Systems 8, 15 and 17). However, as shown by the comparison between System 14 and System 15, combining different augmentation methods does not enhance performance as it does in the pretrained setup. We speculate that optimizing TSE with multiple augmentations is challenging for fully learnable modules.
    \item To evaluate the augmentation methods on a larger dataset, we trained models on the train-460 set. To conserve computational resources, we utilized the pretrained speaker encoder. By comparing System 16 and System 17, we observe that even with a larger training dataset, our methods still yield improvements. Additionally, our approach outperforms the state-of-the-art (SOTA) model from the Espnet recipe.
\end{enumerate}

%  Hence, we design System 16 and System 17, which first pretrained with simple augmentation settings for the first 50 training epoches and then fine tune with the integrated augmentation approach for the next 50 training epoches. 
% `initM' means the model is pretrained from the $50_{th}$ checkpoint of the model without augmentations,  System 15, and `init'  means we load the  $50_{th}$ checkpoint of System 9.  But results show System 16 and 17 don not get obvious improvemnt. In addition, to test our method on large dataset, we design System 18 and 19. To save computational resources, we utilize the pretrained speaker encoder. Results show that even if with large train data, our methods are still useful to get a little improvement.  And it is superior than the SOTA model from Espnet recipe. 

 \vspace{-3mm}
\subsection{Is the improvement from a better speaker encoder?}

Table \ref{tab:noise} demonstrates the effectiveness of augmentation methods. Augmentation improves the robustness of the extraction module, as evidenced by its usefulness even when the speaker model is frozen. However, can augmentation enhances the robustness of the speaker encoder during joint training?
Table \ref{tab:EER} presents the performance of the speaker encoder on the Libri2mix test set. Interestingly, there are no discernible correlations between the EER and the separation performance, suggesting that the augmentation methods do not notably improve the performance of the speaker encoder. This observation suggests that, in comparison to the extraction module, the speaker encoder may play a less critical role in the TSE task.
\vspace{-1mm}
 \subsection{Impact of data augmentation on out-domain datasets}
 \label{sec:out-domain}

Table \ref{tab:100_clean} shows the performance on in-domain test sets (Mix\_both and Mix\_clean) and out-domain test sets (WSJ0-2mix and Aishell2Mix). The main obervations are: 

\begin{enumerate}[leftmargin=0.4cm]
    \item Compared to Table \ref{tab:noise}, the augmentation method on the clean dataset appears to be less effective. We speculate that Mix\_clean is an easier dataset, where the TSE model doesn't require as much data to achieve good performance.
    % Consequently, the effectiveness of augmentation may not be as pronounced.
    \item When the speaker encoder is pretrained, data augmentation benefits both in-domain and out-domain test sets. With a learnable speaker encoder, augmentation also improves performance in both cases, but the improvement is much more significant for out-domain performance. This indicates that data augmentation increases robustness in out-domain scenarios. Similar to Conclusion 2 in section \ref{sec:noise}, the jointly optimized speaker encoder outperforms the pretrained one on both in-domain and out-domain datasets.
    % \item The `init' strategy, which initializes the model from a checkpoint without augmentations, similar to \textbf{curriculum learning}, proves useful for improving performance across most evaluation metrics.
    \item The `CL' strategy, which implements curriculum learning \cite{bengio2009curriculum,wang2021survey} by first training without augmentations and then with them, proves effective for improving performance across most evaluation metrics.
    \item Our best results surpass most SOTA models, with only slightly lower performance compared to SSL-TD-SpeakerBeam and MC-SpEx, which demonstrates the effectiveness of our augmentation methods and the BSRNN backbone.
\end{enumerate}

\vspace{-5mm}
 \subsection{Comparative analysis of data augmentation on enrollment speech and speech mixture }

Due to space constrains, in this section, only noise augmentation is applied to compare the effectiveness of augmentation on enrollment speech versus the speech mixture.

% Data augmentation methods can be applied to either the enrollment speech or the speech mixture. In section \ref{sec:out-domain} the integrated augmentation is utilized, but in this section, we only focus on noise augmentation, where noise is added to either the enrollment speech or the speech mixture using the method described in Equation \ref{equ:noise}. Table \ref{tab:kind} presents the performance of models with augmented enrollment speech or augmented speech mixture. 

\noindent \textbf{Noise augmentation on enrollment speech:}  The augmentation does not  show obvious improvement on in-domain datasets. However, it proves useful on out-domain datasets.

\noindent \textbf{Noise augmentation on speech mixture:} It improves the performance on Mix\_both, which is reasonable. However, it does not show obvious effectiveness on Mix\_clean and out-domain datasets, except for some improvements in terms of accuracy when the speaker encoder is learnable.

In conclusion, augmentation on enrollment speech proves more effective on clean out-domain datasets, while augmentation on speech mixture tends to be more beneficial for in-domain datasets containing similar interference signals, such as noise. These two methods can complement each other to some extent.  Although there may be a decline in performance in certain aspects, the overall performance is improved.

\vspace{-3mm}
\section{Conclusions and future work}
\vspace{-1mm}
In this paper, we highlight the importance and effectiveness of augmenting enrollment speech for the TSE task. We introduce various augmentation methods, including classic noise/reverberation addition, specAugment, and a novel self-estimated speech augmentation (SSA). Validated on both pretrained and jointly trained speaker encoder setups, and both the clean and noisy setups of Libri2Mix, the results demonstrate significant benefits of our proposed method, especially in cross-domain tests. In future work, we plan to explore additional augmentation techniques and investigate their integration with existing methods.

% References should be produced using the bibtex program from suitable
% BiBTeX files (here: strings, refs, manuals). The IEEEbib.bst bibliography
% style file from IEEE produces unsorted bibliography list.
% -------------------------------------------------------------------------
\bibliographystyle{IEEEbib}
\bibliography{main}

\end{document}